\documentclass[3p]{elsarticle}

\usepackage{hyperref}

\usepackage{listings}
\usepackage{tabularx}
\usepackage{multirow}
\usepackage[export]{adjustbox}
\usepackage{amsfonts}
\usepackage{tablefootnote}
\usepackage{amsmath}

\journal{Computers \& Geosciences}

\biboptions{authoryear}

\begin{document}

\begin{frontmatter}

\title{PyRQA - Conducting Recurrence Quantification Analysis on Very Long Time Series Efficiently}

\address[mymainaddress]{Humboldt Universit{\"a}t zu Berlin}
\address[mysecondaryaddress]{GFZ German Research Centre for Geosciences, Potsdam}
\address[mytertiaryaddress]{Potsdam Institute for Climate Impact Research, Potsdam}

\author[mymainaddress,mysecondaryaddress]{Tobias Rawald\corref{mycorrespondingauthor}}
\cortext[mycorrespondingauthor]{Corresponding author}
\ead{tobias.rawald@gfz-potsdam.de}

\author[mysecondaryaddress]{Mike Sips}
\ead{mike.sips@gfz-potsdam.de}

\author[mytertiaryaddress]{Norbert Marwan}
\ead{marwan@pik-potsdam.de}

\begin{abstract}

\emph{PyRQA} is a software package that efficiently conducts recurrence quantification analysis (RQA) on time series consisting of more than one million data points. RQA is a method from non-linear time series analysis that quantifies the recurrent behaviour of systems. Existing implementations to RQA are not capable of analysing such very long time series at all or require large amounts of time to calculate the quantitative measures. \emph{PyRQA} overcomes their limitations by conducting the RQA computations in a highly parallel manner. Building on the OpenCL framework, \emph{PyRQA} leverages the computing capabilities of a variety of parallel hardware architectures, such as GPUs. The underlying computing approach partitions the RQA computations and enables to employ multiple compute devices at the same time. The goal of this publication is to demonstrate the features and the runtime efficiency of \emph{PyRQA}. For this purpose we employ a real-world example, comparing the dynamics of two climatological time series, and a synthetic example, reducing the runtime regarding the analysis of a series consisting of over one million data points from almost eight hours using state-of-the-art RQA software to roughly 69 seconds using \emph{PyRQA}.
\end{abstract}


\end{frontmatter}

\section{Introduction}
\label{introduction}

\emph{Recurrence quantification analysis} (RQA) is a method from nonlinear time series analysis to quantify the recurrent behaviour of systems~\citep{marwan2007}. It has been successfully applied to characterise earthquake dynamics~\citep{chelidze2015}, spatially extended ecosystems~\citep{proulx2009, li2008}, or climate variability~\citep{zhao2011}, to identify nonlinear regimes and complex synchronization in solar variability~\citep{kurths94, zolotova2009}, to investigate past climate teleconnections~\citep{marwan2003climdyn} and regime transitions~\citep{donges2011d}.

RQA relies on the identification of line structures within \emph{recurrence matrices}. Such a matrix is constructed by computing the mutual similarities of multi-dimensional vectors reconstructed from a given time series. Pairs of vectors are either considered similar or dissimilar, resulting in a binary decision. Matrix elements of the same value form line structures. Based on frequency distributions of line lengths, quantitative measures such as the \emph{average line length} are derived.


The underlying algorithms of RQA have a time complexity of $\mathcal{O}(N^2)$. Persisting the binary similarity matrix in the main memory of a compute system results in a quadratic space complexity as well. These properties hamper an efficient analysis of time series consisting more than one million data points. Although short time series are often typical in Geoscience, their size will grow strongly in the future by the increasing effort and success of collecting data and increasing time resolution. 

Existing implementations of RQA suffer from one or multiple limitations, which hamper the analysis of very long time series. This includes the \emph{memory limitation} (not being able to store matrices exceeding the size of the main memory), the \emph{device limitation} (not being able to employ more than one compute device), and the \emph{runtime limitation} (not being able to conduct the analysis in a reasonable amount of time).

\emph{PyRQA} addresses those limitations by introducing concepts from parallel and distributed computing. The underlying computing approach subdivides the binary similarity matrix into multiple sub matrices. Each sub matrix may be processed by a specific compute device in a massively parallel manner. Relying on the \emph{OpenCL} framework, a variety of different hardware architectures, e.g., \emph{graphics processing units} (GPUs), can be employed for processing. Furthermore, the computing approach enables to process multiple sub matrices by multiple compute devices at the same time. As will be shown in Sect.~\ref{sec:synthetic_data}, those properties allow to reduce the runtime of processing a series consisting of over one million data points from almost eight hours to 69 seconds.

\emph{PyRQA} is a free and open-source software package~\citep{pyrqa}. Its computing approach has been successfully applied to a real-world time series from climate impact research that consists of more than one million data points~\citep{Rawald2014}. Here, the runtime of conducting the analysis could be drastically reduced by performing the computations on multiple GPUs. In \cite{Rawald15} it is evaluated, how the application of concepts from database engineering influence the performance of RQA processing. In the following, we focus on the features of \emph{PyRQA} and the usage of its \emph{application programming interface} (API).

The manuscript is structured as follows. Sect.~\ref{basic_concepts} gives a short introduction to the basic concepts of RQA. Sect.~\ref{sec:state_of_the_art_software} describes state-of-the-art implementations of RQA, focussing on their features and limitations. In Sect.~\ref{computing_approach}, the computing approach employed by \emph{PyRQA} is examined in detail. Sect.~\ref{sec:source_code_and_license} presents information on how to obtain the source code of \emph{PyRQA} and the license under which it is released. In Sect.~\ref{installation}, the installation process is described. Sect.~\ref{api_description} outlines the analysis workflow induced by \emph{PyRQA}. In Sect.~\ref{comprehensive_examples}, comprehensive examples on how to employ \emph{PyRQA} are presented. In Sect.~\ref{summary}, a conclusion and an outlook on the future development of \emph{PyRQA} is given.

\section{Basic Concepts of RQA}
\label{basic_concepts}

Environmental systems such as the Earth's climate system show dynamic behaviour, which is highly nonlinear. This behaviour can be observed by monitoring variables, e.g., the air temperature at a specific location. Recurrence analysis is applied, to identify patterns within the temporal dynamics of an environmental system. We refer to~\citep{marwan2007} for more detailed information on the topic of recurrence analysis. 

Recurrence analysis comprises several methods, including the \emph{recurrence plot}. It is based on the reconstruction of multi-dimensional vectors from a given time series by using time-delay embedding \cite{packard80}. Those vectors represent the states of the system under investigation. The pairwise similarity regarding all pairs of vectors reconstructed is determined by applying measures, such as the Euclidean metric. Two vectors are either considered to be similar or dissimilar based on a neighbourhood condition, e.g., a fixed similarity threshold. The corresponding results are captured within the \emph{recurrence matrix}, which has a quadratic shape.

Matrix elements representing pairs of similar vectors are referred to as \emph{recurrence points}, implying that the system recurs to a similar state. A recurrence plot is the visual representation of a recurrence matrix, encoding recurrence points as black dots and the remaining elements as white dots. These dots form small-scale structures, in particular lines. Visually inspecting those structures allows to draw conclusions regarding the temporal dynamics of a system. 

A recurrence plot is only applicable for time series consisting of hundreds of data points, due to screen size limitations. Only parts of the plot can be displayed, if the number of data points is increased to hundreds of thousands. Furthermore, applying downsampling strategies may create artifacts within a recurrence plot, causing false interpretations~\citep{marwan2011}. Referring to the information captured within a recurrence plot, recurrence quantification analysis is a method to quantitatively assess its visual impression~\citep{zbilut2007a}. This quantification was later connected to theoretical understanding~\citep{marwan2007}.

RQA quantifies the line structures within the recurrence plot consisting of recurrence and non-recurrence points. It considers three different types of lines, each of them assigned with specific semantics:

\begin{enumerate}

\item diagonal lines (recurrence points), 
\item vertical lines (recurrence points), and
\item white vertical lines (non-recurrence points).

\end{enumerate}

\begin{figure}
  \centering
  \includegraphics[width=0.8\columnwidth]{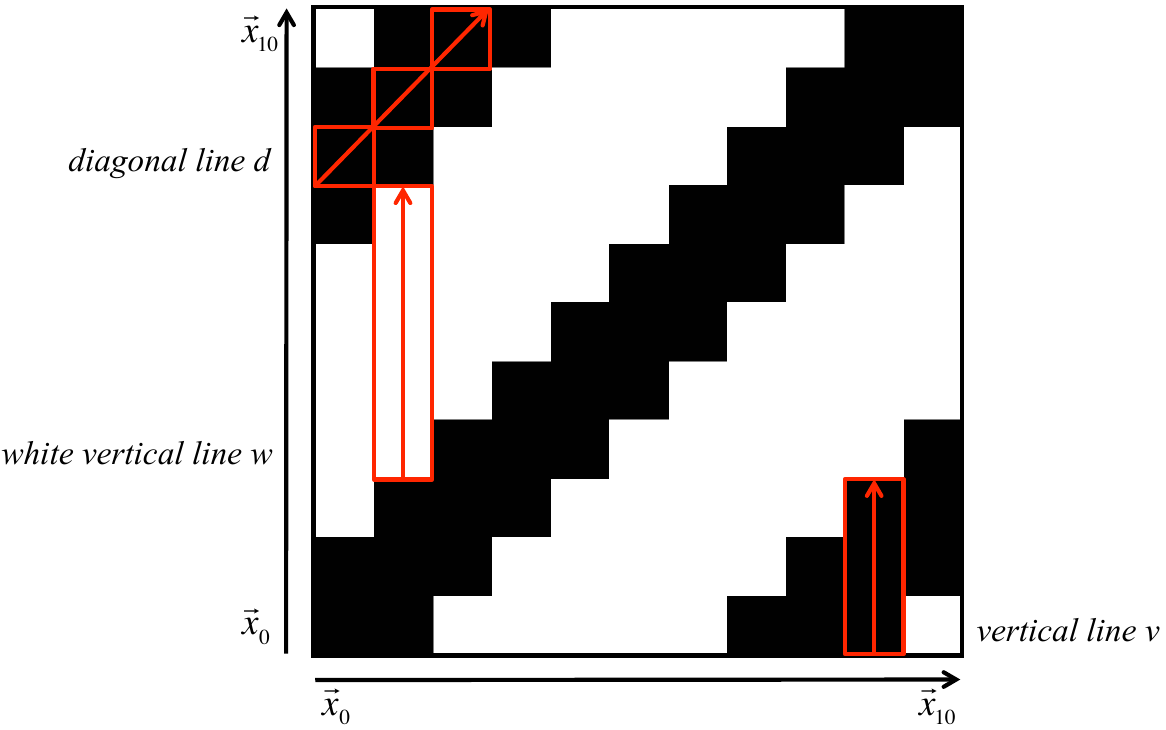}
  \caption{Basic RQA Concepts. The recurrence plot displayed is based on a time series created by observing an oscillating system. Eleven vectors, $\vec{x}_{0}$ to $\vec{x}_{10}$, are reconstructed from the series. A diagonal line $d$ of length $3$, a vertical line $v$ of length $3$ and a white vertical line $w$ of length $5$ are highlighted.}
  \label{fig:basic_concepts}
\end{figure}

For each of those types, frequency distributions of their occurrences are computed. Based on those distributions, quantitative measures are derived, e.g., the portion of recurrence points that form diagonal lines (\emph{determinism}, $DET$). The set of quantitative measures describes the temporal dynamics of a system under investigation. Unless stated otherwise, in the following the expression ``diagonal and vertical lines'' refers to all three line types. The basic concepts behind RQA are depicted in Fig.~\ref{fig:basic_concepts} and the definition of the RQA measures can be found in \citet{marwan2007}.

\section{State-of-the-Art in RQA Software}
\label{sec:state_of_the_art_software}

In the following, an overview of software that allows to conduct RQA or create recurrence plots is given\footnote{A continuously maintained list of software is available at \url{http://www.recurrence-plot.tk/programmes.php}.}. The focus lies on implementations that are freely available as well as open source and do not rely on proprietary software, such as MATLAB. Tools are distinguished regarding their functionality and their computational limitations.

The RQA software presented in the following suffers from multiple of the following limitations regarding the analysis of very long time series, consisting of hundreds of thousands of data points. Table~\ref{tab:software_limitations} maps each software tool to its limitations.

\begin{description}
	\item[Memory limitation:] The recurrence matrix does not fit into the main memory of the computing system. Matrices exceeding the size of the main memory can not be analysed as a whole or can not be analysed at all.
	\item[Device limitation:] The quantitative analysis is conducted on a single CPU. The computational capabilities of systems containing multiple CPUs or accelerators, such as GPUs, are not exploited.
	\item[Runtime limitation:] The quantitative analysis consumes large amounts of time, due to the limited exploitation of parallel processing capabilities. 
\end{description}

\begin{table*}
\caption{Limitations of Existing RQA Software. Each software tool is mapped to its limitations. The tools that suffer from memory limitation do not necessarily consume large amounts of runtime, since the size of recurrence matrices is restricted, such that they are not capable of analysing very long time series as a whole. Nonetheless, they only make limited use of parallel processing techniques. Note, the tool \texttt{recurr} that is part of \emph{TISEAN} is only capable of creating recurrence plots.}\label{tab:software_limitations}
\begin{center}
\begin{tabular}{|l|p{.15\textwidth}|p{.15\textwidth}|p{.15\textwidth}|}
\hline
\emph{Software Tool}				& \emph{Memory Limitation}	& \emph{Device Limitation}	& \emph{Runtime Limitation}	\\
\hline
\hline
\emph{RQA X}						& $\checkmark$			& $\checkmark$			& ($\checkmark$)			\\
\hline
\emph{TISEAN} (\texttt{recurr})			&						& $\checkmark$			& $\checkmark$			\\
\hline
\emph{crqa}						& $\checkmark$			& $\checkmark$			& ($\checkmark$)			\\
\hline
\emph{pyunicorn}					& $\checkmark$			& $\checkmark$			& ($\checkmark$)			\\
\hline
\emph{Commandline Recurrence Plots}	&						& $\checkmark$			& $\checkmark$			\\ 
\hline
\end{tabular}
\end{center}
\end{table*}

\subsection{RQA X}

\emph{RQA X}~\citep{Keller2016} supports the construction of recurrence matrices based on a fixed radius neighbourhood by employing a variety of similarity measures, such as the $L_{2}$-norm and the $L_{\infty}$-norm, to conduct the pairwise comparison of vectors. Regarding the quantitative analysis, RQA X does not provide any information regarding white vertical lines. The software is written in \emph{Objective C} and provides similar functionality to \emph{RQA Software}~\citep{Webber2016}.

RQA X is only capable of analysing recurrence matrices created from up to 40,000 vectors\footnote{This value is hardcoded in the file \texttt{RQAPrefsController.m} as \texttt{maximumBatchWindowSize}.}. To compensate this restriction, it is possible to specify \emph{epochs}, which are fixed-sized windows along the middle diagonal of the full recurrence matrix~\citep[p. 52]{Webber2005b}.

RQA X performs its computations solely on CPU devices. It allows to conduct multiple quantitative analyses at the same time; each analysis is conducted within a separate CPU thread. 
	
\subsection{TISEAN}

\emph{TISEAN}, an acronym derived from \emph{time series analysis}, is a collection of command line tools released under \emph{GPL} license, which allow to analyse numerical time series. Version 3.0.1 comprises utilities, e.g., for generating time series or performing noise reduction as well as conducting linear and nonlinear time series analysis~\citep{Hegger1999}. 

TISEAN includes two versions of the program \texttt{recurr}, one written in C and one written in FORTRAN, to compute the content of recurrence matrices. \texttt{recurr} computes the pairwise distances of the reconstructed vectors using the fixed radius neighbourhood in combination with the $L_{\infty}$-norm. The output of \texttt{recurr} is a list of recurrence points, represented as pairs of integer values that may either be written to \emph{stdout} or stored in a file. \texttt{recurr} does not provide any functionality to conduct RQA based on those recurrence points.

\subsection{crqa}

\emph{crqa} is a package written in \emph{R} and released under \emph{GPL} license. It allows to conduct \emph{cross recurrence quantification analysis} (cRQA)~\citep{Coco2014}. The R package is partially based on the \emph{Cross Recurrence Plot Toolbox} that is implemented in MATLAB~\citep{CRPToolbox}. The method cRQA differs from traditional RQA by comparing the recurrent behaviour captured in two time series. The recurrence vectors reconstructed from each time series are assigned to one of the two axis of the recurrence matrix. The determination of the pairwise similarities as well as the detection of line structures is similar.


Among others, the package crqa provides the method \verb+crqa+, which performs cRQA on two input time series. It can imitate traditional RQA by providing the same input time series twice. \verb+crqa+ computes the recurrence plot as well as nine quantitative measures, e.g., \emph{recurrence rate} and \emph{determinism}. The method constructs a cross recurrence matrix based on a fixed radius neighbourhood. 

The data structures employed by \verb+crqa+ persist the recurrence matrix in the main memory of the computing system. Hence, it is only capable of analysing recurrence matrices that fit into its memory, limiting the length of the time series to investigate to a couple of thousands. \verb+crqa+ uses parallelisation techniques by executing multiple CPU threads while constructing and analysing the recurrence matrix. 

\subsection{pyunicorn}

\emph{pyunicorn} is a Python package released under \emph{BSD} license that aims at conducting complex network as well as recurrence analysis~\citep{Donges2015}. Amongst others, it allows to perform RQA and to create recurrence plots. To improve the efficiency of the computations, it includes code fragments that are written in C, C++ and FORTRAN.

pyunicorn persists the recurrence matrix within the main memory during the detection of diagonal and vertical lines. Depending on the size of the main memory available, this property limits the length of the time series that can be processed to only a couple of thousands of data points. This renders the analysis time series consisting of hundreds of thousands of data points impossible.

\subsection{Commandline Recurrence Plots}

\emph{Commandline Recurrence Plots} allows to compute recurrence plots and to conduct recurrence quantification analysis~\citep{CommandlineRecurrencePlots}. Version 1.13z of the tool can be obtained in compiled form for a variety of platforms, including Linux, Mac OS X, Windows, HP-UX and Solaris. Note, its source code is not publicly available.


The focus of Commandline Recurrence Plots is conducting RQA. The identification of line structures is performed without storing the recurrence matrix in the memory of the computing system. The similarity values referring to pairs of vectors are rather computed on the fly, while sequentially inspecting the elements within diagonals and columns of the matrix. This allows to analyse recurrence matrices of almost arbitrary size. 

Commandline Recurrence Plots conducts the computations solely in a single CPU thread. This does only employ a fraction of the computing capabilities provided by multi-core CPUs.

\section{Distributed and Parallel Computing Approach}
\label{computing_approach}

%
%


We present insights on the underlying computing approach of \emph{PyRQA}. The focus lies on the RQA processing, although it also support the creation of recurrence plots. \emph{PyRQA} overcomes the limitations of existing RQA software, as described in Sect.~\ref{sec:state_of_the_art_software}, by:

\begin{itemize}
	\item subdividing the full recurrence matrix into multiple sub matrices that fit into the memory,
	\item distributing the processing of the sub matrices across multiple compute devices, and
	\item conducting the computations within a single sub matrix in a massively parallel manner.    
\end{itemize}

In the following, those aspects are explained in detail.

\subsection{Divide \& Recombine}

Recurrence matrices exceeding the size of the memory available cannot be stored in their entirety. The computing approach of \emph{PyRQA} applies the concept \emph{Divide \& Recombine}~\citep{guha2012} to enable their processing. The recurrence matrix is subdivided into a set of sub matrices. The calculation of the pairwise vector similarities as well as the detection of line structures is performed for each sub matrix individually. The separate sub matrix results are recombined into global data structures, which serve as the basis for computing the RQA measures. 

Lines may cross the vertical and horizontal borders of adjacent sub matrices. \emph{PyRQA} employs additional data structures, to ensure their correct detection. The \emph{carryover buffers} store the length of lines that reach the outer borders of the sub matrix currently inspected. An individual carryover buffer is provided for each line type. Those intermediate line lengths are used as an input for the line detection in adjacent sub matrices. The overhead for storing and maintaining those carryover buffers is marginal.

Dividing the full recurrence matrix has two major benefits. First, it overcomes the \emph{memory limitation}, since the size of the sub matrices can be chosen such that they fit into the memory space available. Second, the processing of sub matrices can be distributed across multiple compute devices, thus there is no \emph{device limitation}. 


More information on the functionality of the carryover buffers as well as the properties of the sub matrix processing order can be found in~\cite{Rawald2014}.

\subsection{Massively Parallel Sub Matrix Processing}

The software tools presented in Sect.~\ref{sec:state_of_the_art_software} only use the parallel processing capabilities provided by a CPU or none at all. This neglects the fact that the quantitative analysis can be subdivided into a set of distinct operators, where each operator itself can be processed in a massively parallel manner. Those operators are applied to each sub matrix individually.

The computing approach of \emph{PyRQA} distinguishes between the operators:

\begin{enumerate}[I]
	\item \emph{create\_recurrence\_matrix},
	\item \emph{detect\_diagonal\_lines}, and
	\item \emph{detect\_vertical\_lines}.
\end{enumerate}

There is the constraint that the \emph{create\_recurrence\_matrix} operator has to be executed before the line detection operators are started. Each operator can be mapped to a type of atomic tasks:

\begin{enumerate}[I]
	\item the similarity comparison of a single pair of reconstructed vectors,
	\item the inspection of a single diagonal of the recurrence matrix regarding diagonal lines, and
	\item the inspection of a single column of the recurrence matrix regarding vertical and white vertical lines.
\end{enumerate}

Each atomic task is fully independent of any other task of the same operator. This property allows to  execute multiple tasks of the same operator in parallel. Each operator is assigned with a maximum degree of parallelism ($DOP_{max}$), based on the number of reconstructed vectors $N$:

\begin{enumerate}[I]
	\item  $DOP_{max} = N^2$,
	\item  $DOP_{max} = 2N - 1$ (non-symmetric recurrence matrix) / $DOP_{max} = N - 1$ (symmetric recurrence matrix), and
	\item  $DOP_{max} = N$.
\end{enumerate}

The maximum DOP captures the maximum number of tasks that can run in parallel per operator. This parallel execution contributes to overcoming the \emph{runtime limitation} by maximising the utilisation of the computing resources of a single compute device.

\subsection{Technical Details}
\label{technical_details}

\emph{PyRQA} employs the OpenCL framework for heterogeneous computing~\citep{StoneGS10}. OpenCL is designed to exploit the parallel computing capabilities of multi-core devices, such as CPUs, and many-core devices, such as GPUs. We have chosen OpenCL, because it is supported by a variety of compute devices from different hardware vendors.

The processing of OpenCL is subdivided between a single \emph{host device} and one or more \emph{compute devices}. The source code of \emph{PyRQA} consists of a \emph{host program} written in \emph{Python} and \emph{kernel functions} written in \emph{OpenCL C}. The kernel functions capture the atomic tasks of each operator and are executed by the compute devices. 

Detailed information regarding the performance of a set of RQA implementations using OpenCL running on compute devices from different hardware vendors is provided by~\cite{Rawald15}. 

\section{Source Code and License}
\label{sec:source_code_and_license}

\emph{PyRQA} is distributed via the \emph{Python Package Index}~\citep{pyrqa}. Its contents are free of charge as well as open source and released under version 2.0 of the \emph{Apache License}. The source files of version 0.1.0 can be downloaded at~\cite{pyrqa_source}. It is planned to open the existing GitLab repository for public access in the near future~\citep{pyrqa_gitlab}.
\section{Installation}
\label{installation}


\emph{PyRQA} can be installed using the command line tool \emph{pip}~\citep{pip}, using the command:

\begin{verbatim}

pip install PyRQA

\end{verbatim}

The package defines dependencies regarding additional Python packages that are required for execution, including \mbox{\emph{NumPy}}~\citep{numpy}, \mbox{\emph{PyOpenCL}}~\citep{pyopencl}, and \mbox{\emph{Pillow}}~\citep{pillow}. Using \emph{pip}, those dependencies are installed automatically. 

\emph{PyRQA} may require the installation of OpenCL related software, such as custom hardware drivers and the OpenCL runtime. The amount of software to be installed is device-specific and varies between hardware vendors:

\begin{description}
\item[AMD:] \url{http://developer.amd.com/tools-and-sdks/opencl-zone}
\item[Intel:] \url{http://software.intel.com/en-us/articles/opencl-drivers}
\item[Nvidia:] \url{http://developer.nvidia.com/opencl}
\end{description}

Apple as a key driver of OpenCL provides support within the latest versions of \emph{Mac OS X} by default. If \emph{PyRQA} is unable to detect any OpenCL compute device, exceptions are thrown during its execution.
\section{PyRQA Analysis Workflow}
\label{api_description}

This section describes the typical workflow of using \emph{PyRQA} as well as the relevant package contents. An analysis is structured into four processing steps:

\begin{enumerate}
	\item Retrieving a time series to investigate,
	\item Assigning values to the analysis parameters,
	\item Creating the analysis computation, and
	\item Retrieving the final computing results.
\end{enumerate}

Those steps are similar for conducting RQA as well as creating a recurrence plot. Note, the example source code displayed in the following refers to a RQA scenario.

Adhering to the object oriented programming approach, \emph{PyRQA} comprises software components that encapsulate the functionality employed within each processing step. A comprehensive documentation of the \emph{PyRQA} API is provided by~\cite{pyrqa_documentation}. An overview of the object classes mentioned in following is given in Tab.~\ref{tab:class_overview}.

\begin{table*}
\caption{Overview of Object Classes. For each processing step, the object classes employed within the following examples are captured. For each object class, a short description is provided. A complete documentation of all object classes of \emph{PyRQA} is given by~\cite{pyrqa_documentation}.}\label{tab:class_overview}
\begin{center}
\begin{tabular}{|c|l|p{0.45\textwidth}|}
\hline
\emph{Step}			& \emph{Object Class}				& \emph{Description}				\\
\hline
\hline
\multirow{1}{*}{1.}		& \verb+FileReader+					& Extract time series from file.			\\
\hline
\multirow{3}{*}{2.}		& \verb+Settings+					& Set of analysis parameters.			\\
					& \verb+EuclideanMetric+				& Similarity measure.				\\
					& \verb+FixedRadius+				& Neighbourhood condition.			\\
\hline
\multirow{3}{*}{3.}		& \verb+RQAComputation+			& RQA computation.					\\
					& \verb+RecurrencePlotComputation+	& Recurrence plot computation.			\\
					& \verb+OpenCL+					& OpenCL environment.				\\
\hline
\multirow{2}{*}{4.}		& \verb+RQAResult+					& RQA result.						\\
					& \verb+RecurrencePlotResult+		& Recurrence plot result.				\\
\hline
\end{tabular}
\end{center}
\end{table*}

\subsection{Retrieving a Time Series to Investigate}

%
%
%

A common format to represent time series data are text files containing values separated by delimiters, such as \emph{comma-separated values} (CSV). Each separated column within such a file by convention represents a time series referring to a specific observational variable. The object class \verb+FileReader+ provides means to extract such series from delimiter-separated files.

\begin{lstlisting}[language=Python, 
			caption={Retrieving a Time Series.}, 
			label={lst:retrieving_time_series}, 
			basicstyle=\footnotesize, 
			frame=single]
from pyrqa.file_reader import FileReader

time_series = FileReader.file_as_float_array('csv_data.txt', 
                                             delimiter=',',
                                             column=3, 
                                             offset=10)
\end{lstlisting}

The static method \verb+file_as_float_array+ reads a column from an input file and transforms it into an array of floating point values. A corresponding call is depicted in List.~\ref{lst:retrieving_time_series}. The \verb+delimiter+ refers to a single character that separates the individual columns. The \verb+column+ from which the time series should be extracted is referred to by an integer, starting at zero. It is possible to omit a fixed number of values at the beginning of the column, by specifying an \verb+offset+.

\subsection{Assigning Values to the Analysis Parameters}

RQA and recurrence plot rely on the specification of a set of input parameters. The values of those parameters are encapsulated within an object of the class \verb+Settings+. The creation of such an object is depicted in List.~\ref{lst:assigning_parameters}. The set of parameters include:
 
\begin{itemize}
	\item the input \verb+time_series+,
	\item the \verb+embedding_dimension+ and the \verb+time_delay+ parameter used during the vector reconstruction,
	\item the \verb+similarity_measure+ used to compare the vectors, as well as
	\item the \verb+neighbourhood+ used to detect neighbouring vectors.
\end{itemize}

Version 0.1.0 of \emph{PyRQA} supports the similarity measures $L_{1}$-norm, $L_{2}$-norm and $L_{\infty}$-norm. A fixed radius can be applied as neighbourhood condition. Moreover, RQA relies on the specification of minimum line lengths, including:

\begin{itemize}
	\item the \verb+min_diagonal_line_length+,
	\item the \verb+min_vertical_line_length+, and
	\item the \verb+min_white_vertical_line_length+.
\end{itemize}


\begin{lstlisting}[language=Python, 
			caption={Assigning Values to the Analysis Parameters.}, 
			label={lst:assigning_parameters}, 
			basicstyle=\footnotesize, 
			frame=single]
from pyrqa.settings import Settings
from pyrqa.metric import EuclideanMetric
from pyrqa.neighbourhood import FixedRadius

settings = Settings(time_series,
                    embedding_dimension=2,
                    time_delay=3,
                    similarity_measure=EuclideanMetric,
                    neighbourhood=FixedRadius(radius=2.0),
                    min_diagonal_line_length=2,
                    min_vertical_line_length=2,
                    min_white_vertical_line_length=2)
\end{lstlisting}

\subsection{Creating the Analysis Computation}

\emph{PyRQA} comprises two object classes as an abstraction layer to create a computation based on the settings and other arguments specified; \verb+RQAComputation+ and \verb+RecrurrencePlotComputation+. This is realised using the static method \verb+create+ that is provided by both object classes.

The only required argument of the \verb+create+ method is the \verb+Settings+ object, capturing the analysis parameters. An optional argument is the OpenCL environment, which is encapsulated within an object of the class \verb+OpenCL+ (see List.~\ref{lst:handing_over_opencl}). The OpenCL environment is determined automatically, if the \verb+opencl+ keyword argument is not assigned.

On its creation, an \verb+OpenCL+ object discovers the OpenCL platforms as well as compute devices available in the computing system. There are two ways to determine the OpenCL environment. First, platform and device selection can be performed using the method \verb+create_environment+. This method employs the \verb+platform_id+ and the list of \verb+device_ids+ that are given as parameters to the constructor of the \verb+OpenCL+ object. Those IDs can be determined by a prior inspection of the platforms and devices available. Second, the discovery can be performed using the method \verb+create_environment_command_line+. Here, the selection is conducted by command line input. The method called depends on the value of the constructor parameter \verb+command_line+, which is either set \verb+True+ or \verb+False+.

\begin{lstlisting}[language=Python, 
			caption={Creating the Analysis Computation.}, 
			label={lst:handing_over_opencl}, 
			basicstyle=\footnotesize, 
			frame=single]
from pyrqa.opencl import OpenCL
from pyrqa.computation import RQAComputation

opencl = OpenCL(platform_id=0,
		device_ids=(0,)
		command_line=True,
		optimisations_enabled=False)
            
computation = RQAComputation.create(settings,
                                    opencl=opencl)
\end{lstlisting}

The \verb+OpenCL+ object is further responsible for compiling the kernel functions written in OpenCL C. In this regard, the compiler can be advised to activate certain optimisations, such as relaxed math operations. Several optimisations are activated by default. Those \emph{default optimisations} are platform and device specific. To avoid inconsistency regarding the computing results across multiple plaforms, it might be required to deactivate those default optimisations. For this reason, the constructor of the \verb+OpenCL+ object provides the flag \verb+optimisations_enabled+, which is set to \verb+False+ by default.

\subsection{Retrieving the Final Computing Results}

Having created the computation object, the execution is started by calling its method \verb+run+, which controls the processing of the analysis operators. \verb+run+ returns the final computing results, which are encapsulated in an object of the class \verb+RQAResult+ or \verb+RecurrencePlotResult+, depending on the analysis method chosen. Conducting RQA, the corresponding result object contains a value for each quantitative measure. Those measures are members of the \verb+RQAResult+ object, e.g. \verb+determinism+ (see List.~\ref{lst:retrieving_computing_results}).

\begin{lstlisting}[language=Python, 
			caption={Retrieving the Final Computing Results.}, 
			label={lst:retrieving_computing_results}, 
			basicstyle=\footnotesize, 
			frame=single]
result = computation.run()
print result.determinism
\end{lstlisting}
\section{Application Examples}
\label{comprehensive_examples}


This section presents two examples regarding the application of \emph{PyRQA}, focussing explicitly on the capability, reproducibility, and performance of the implementation and computing results, respectively. The first example refers to real-world climatological data that is freely available and addresses the capability of the method itself and the qualitative reproducibility of the RQA results. The second example employs a synthetic series that exceeds one million data points. It highlights the performance improvements gained by using \emph{PyRQA} in combination with parallel computing hardware. The API used refers to version 0.1.0 of \emph{PyRQA}.


\subsection{Climatological Data}

In this example we compare the recurrence properties of air temperature during a winter and a summer month. The recurrence properties should allow to compare the temperature dynamics between summer and winter season, e.g., whether the temperature would evolve in a more erratic or more regular way.
For this analysis we have selected the measuring station at the \emph{Asheville Regional Airport} in North Carolina, those data is provided by the \emph{National Oceanic and Atmospheric Administration} (NOAA) of the United States of America~\citep{qclcd} as \emph{Quality Controlled Local Climatological Data} (QCLCD). Regarding the observational variable, the hourly dry-bulb temperature in degree Celsius (column: \verb+DryBulbCelsius+) of January and July 2016 is investigated. Only those measurements are considered that refer to minute 54 of each hour. This results in two time series consisting of 744 data points.

\begin{figure}[htbp]
  \centering
  \includegraphics[width=\columnwidth]{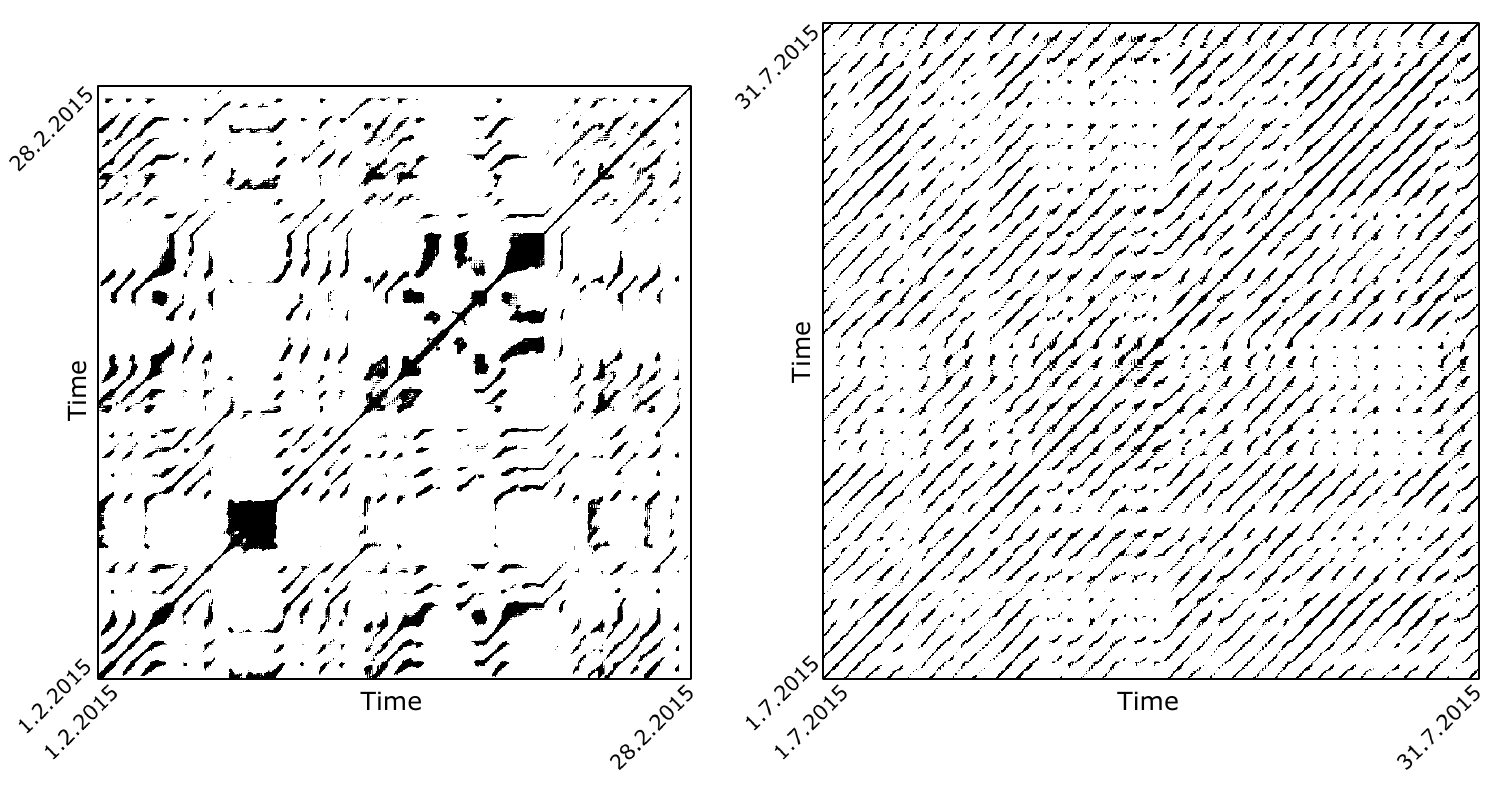}
  \caption{Recurrence Plots of the Hourly Dry-bulb Temperature at Ashville Regional Airport. The plot referring to January 2016 (left) is compared to the plot referring to July 2016 (right).}
  \label{fig:real_world_example_rp}
\end{figure}

The measurement data can be obtained by querying the corresponding web form (selection: \verb+Hourly (10A)+). The query returns markup files that contain comma-separated data and is processed as presented in~\ref{real_world_example}. Similar parameter assignments for reconstructing the system states as in~\cite{Rawald2014} are used. The recurrence threshold has been selected in such a way that the recurrence rate in both examples is roughly $10\%$ \citep{Schinkel2008}. The corresponding recurrence plots are shown in Fig.~\ref{fig:real_world_example_rp} and the RQA results are presented in Tab.~\ref{tab:real_world_example_rqa}. The source code for performing the quantitative analysis as well as creating the corresponding recurrence plots is presented in~\ref{real_world_example}. In order to check the reproducibility we have repeated the RQA by using the {\it Commandline Recurrence Plots} software.


The recurrence plots of January and July exhibit different appearances (Fig.~\ref{fig:real_world_example_rp}); where in July the recurrence plot consists of many diagonal lines indicating clearly the daily variation, the January recurrence plot consists of fewer diagonal lines and empty regions that express a less stationary dynamics during this season. Most RQA measures quantify these differences; only the measures $RR$, $L_\text{max}$, and $DIV$ are equal (Tab.~\ref{tab:real_world_example_rqa}). For January we find higher values in $DET$ and $L$ than for July, but this difference comes from increased occurrences of extended black regions in the January recurrence plot (Fig.~\ref{fig:real_world_example_rp}, left), what is confirmed by the elevated values of $LAM$, $TT$, and $V_{\text{max}}$ for January. Higher values in the entropy measures $L_{\text{entr}}$ and $V_{\text{entr}}$ reveal the more complex distribution of diagonal and vertical lines, as it is also visible in the recurrence plot by the extended white regions and interrupted diagonal lines for January. 
All these quantitative results allow us to conclude that during the summer months the temperature dynamics evolve more periodic and less irregular than during the winter months. (Note, these tentative results and conclusions are derived from just one year and would need deeper investigation and statistical justification that is outside the scope of this paper. The focus of this application example is rather to demonstrate the features of \emph{PyRQA} based on a real-world example using time series that are freely available.)

\begin{table}[htbp]
\caption{RQA Results of the Hourly Dry-bulb Temperature at Ashville Regional Airport for January and July 2016 using the \emph{PyRQA} software. Results from the {\it Commandline Recurrence Plots} software are parenthesized and are in agreement with the \emph{PyRQA} results	. 
}\label{tab:real_world_example_rqa}
\begin{center}
\begin{tabular}{|l|r|r|}
\hline
\emph{RQA Measure}						& \emph{Jan 2016}	& \emph{Jul 2016}	\\
\hline
\hline
Recurrence rate ($RR$)						& 0.10 (0.10)			& 0.10 (0.10)			\\
\hline
Determinism ($DET$)						& 0.94 (0.94)			& 0.88 (0.88)			\\
\hline
Average diagonal line length ($L$)				& 7.80 (7.80)		& 5.75 (5.75)			\\
\hline
Longest diagonal line length ($L_{\text{max}}$)			& 732 (732)			& 732 (732)			\\
\hline
Divergence ($DIV$)							& 0.001 (0.001)		& 0.001 (0.001)			\\
\hline
Entropy diagonal lines ($L_{\text{entr}}$)				& 2.67 (2.67)			& 2.14 (2.14)			\\
\hline
Laminarity ($LAM$)							& 0.97 (0.97)			& 0.94 (0.94)			\\
\hline
Trapping time ($TT$)						& 7.01 (7.01)			& 3.63 (3.63)			\\
\hline
Longest vertical line length ($V_{\text{max}}$)			& 62	 (62)			& 13	(13)			\\
\hline
Entropy vertical lines ($V_{\text{entr}}$)				& 2.62 (2.62)			& 1.65 (1.65)			\\
\hline

\end{tabular}
\end{center}
\end{table}

For the reproducibility test we have recalculated the RQA measures from the two temperature time series applying the {\it Commandline Recurrence Plots} software and using the exact same parameters as for the \emph{PyRQA}. We found exactly the same results (Tab.~\ref{tab:real_world_example_rqa}), confirming the validness of the \emph{PyRQA} implementation and the reproducibility of its results.

\subsection{Synthetic Data}
\label{sec:synthetic_data}


This example demonstrates the performance improvements of employing \emph{PyRQA} in combination with parallel computing hardware, in particular multi-core CPUs and many-core GPUs. The runtime is compared to version 1.13z of the tool Commandline Recurrence Plots. Note, the other tools presented in Sect.~\ref{sec:state_of_the_art_software} are not capable of conducting RQA on such very long time series at all or allow only to analyse parts of it.

For this purpose of reproducibility, a synthetic series based on the sine function consisting of 1,000,001 data points is generated using the Python package Numpy (see List.~\ref{lst:creation_synthetic_series}). This series serves as an input to the RQA computation. An embedding dimension of $2$, an time delay of $2$ and a fixed radius of $1.0$ are applied. The minimum line lengths are set to $2$.

\begin{lstlisting}[language=Python, 
			caption={Creation of Synthetic Series.}, 
			label={lst:creation_synthetic_series}, 
			basicstyle=\footnotesize, 
			frame=single]
import numpy as np

x = np.linspace(0,
                1000 * np.pi,
                1000001)
time_series = np.sin(x)

\end{lstlisting}

The computing system employed for conducting the runtime experiments contains an Intel Core i7-3820 CPU providing 4 cores and 8 threads. Each core runs at up to 3.8 GHz. The CPU accesses 16 GB of main memory. The computing system further includes two NVIDIA GeForce GTX 690 GPUs. Each GPU is equipped with two graphics processors. Each processor is provided with 2 GB of dedicated memory.

\begin{figure}
  \centering
  \includegraphics[width=\columnwidth]{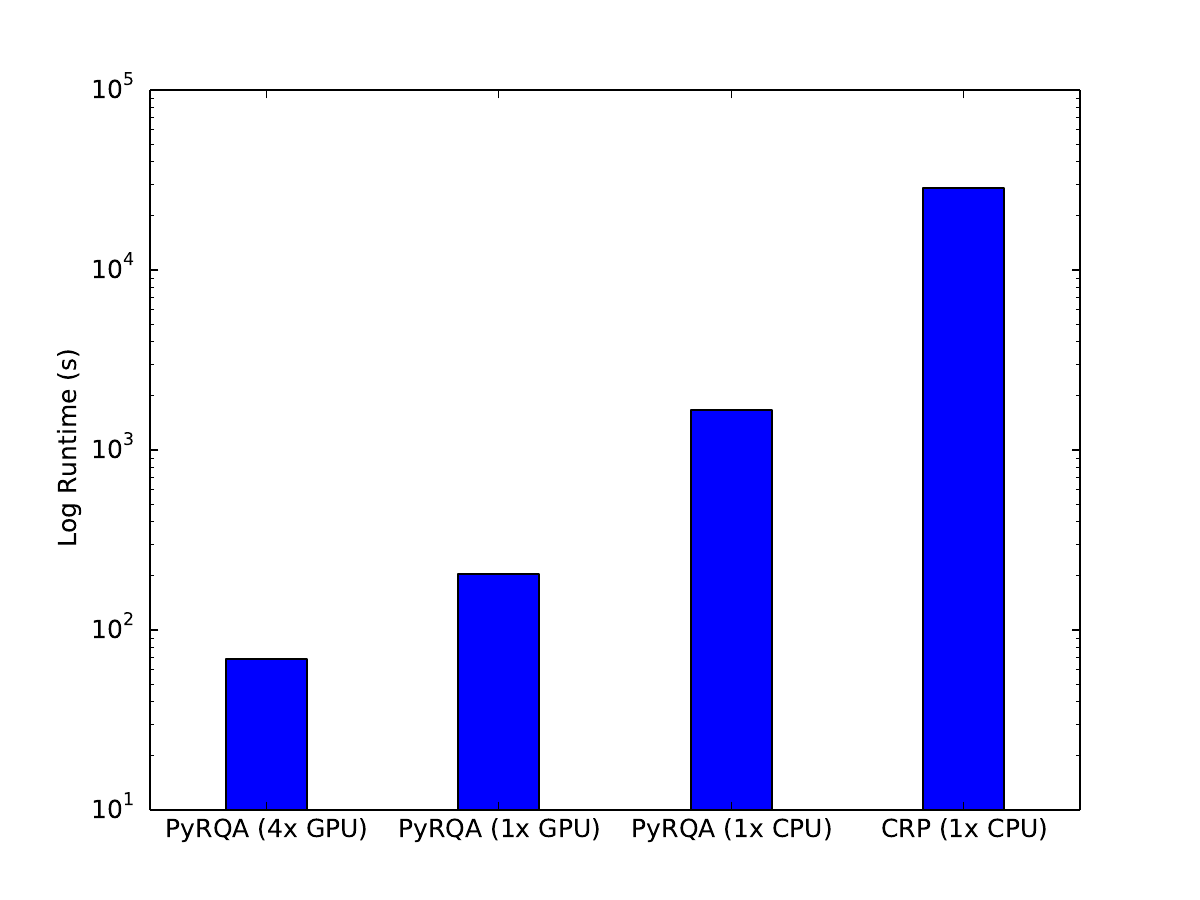}
  \caption{Runtime Results of the Synthetic Application Example. The runtimes are depicted using a logarithmic scale.}
  \label{fig:runtimes_log}
\end{figure}

The total runtimes of executing the following configurations are compared in Tab.~\ref{tab:runtime_results} and Fig.~\ref{fig:runtimes_log}:

\begin{description}
	\item[PyRQA (4x GPU):] \emph{PyRQA} using the four GPU processors of the two NVIDIA GeForce GTX 690.
	\item[PyRQA (1x GPU):] \emph{PyRQA} using a single GPU processor of one NVIDIA GeForce GTX 690. 
	\item[PyRQA (1x CPU):] \emph{PyRQA} using the Intel Core i7-3820 CPU.
	\item[CRP (1x CPU):] Commandline Recurrence Plots using the Intel Core i7-3820 CPU.
\end{description}

\begin{table}
\caption{Runtime Results of the Synthetic Application Example. The runtimes for four different configurations are presented. In addition, for each pair of configuration the corresponding speedup is listed.}
\label{tab:runtime_results}
\begin{center}
\begin{tabularx}{\columnwidth}{|l|r|X|X|X|}
\hline
\emph{Configuration}					& \emph{Runtime (s)}	& \multicolumn{3}{c}{Speedup in Comparison to}	\vline								\\
									&					& \emph{PyRQA} (1x GPU)	& \emph{PyRQA} (1x CPU)	& CRP  \newline (1x CPU)		\\ 
\hline
\hline
\emph{PyRQA} (4x GPU)					& 68.94				& 2.97					& 24.19 					& 413.13					\\
\hline
\emph{PyRQA} (1x GPU)					& 204.81				& -						& 8.14					& 139.06					\\
\hline
\emph{PyRQA} (1x CPU)					& 1,667.58			& -						& -						& 17.08					\\
\hline
CRP (1x CPU)							& 28,480.09			& -						& -						& -						\\
\hline
\end{tabularx}
\end{center}
\end{table}

The OpenCL kernels executed are compiled with default optimisations on each platform. Each total runtime value is computed as the average of four experimental runs. 


Executing the synthetic example on four GPU processors using \emph{PyRQA} has the lowest total runtime, roughly 69 seconds. It delivers a speedup of a factor of almost three in comparison to using only a single GPU processor. The initial overhead of creating the OpenCL environment as well as the final computation of the RQA measures reduce the theoretical speedup of four. The speedup of using four GPUs in comparison to using the Commandline Recurrence Plots tool is more than 400, signalling a drastic performance improvement.

A major benefit of using the OpenCL framework is to leverage the parallel computing capabilities of a variety of devices, including multi-core CPUs. Executing \emph{PyRQA} on the Core i7 CPU delivers a speedup of 17 in comparison to executing Commandline Recurrence Plots on the same device. Among others, this is due to performing the processing within eight CPU threads instead of only one.

\section{Conclusion and Outlook}
\label{summary}

\emph{PyRQA} is a free and open source Python package that allows to efficiently conduct RQA for time series consisting of more than one million data points. Its major contributions are:

\begin{description}
	\item[Distributed and Parallel Computing Approach:] The underlying computing approach of \emph{PyRQA} employs the concept Divide \& Recombine. It subdivides the full recurrence matrix into sub matrices, performs the detection of diagonal and vertical line structures within a sub matrix in a massively parallel manner across multiple compute devices, and recombines the individual sub matrix results into a global result.
	\item[Usage of OpenCL:] \emph{PyRQA} employs the OpenCL framework for heterogenous computing. This allows to conduct RQA in a parallel fashion on a variety of compute devices, including GPUs and CPUs, from different hardware vendors, without having to modify the source code.
\end{description}

The combination of both contributions allows \emph{PyRQA} to overcome the limitations of existing RQA implementations, which are either not capable of performing RQA on very long time series at all or consume large amounts of time. Using a synthetic series consisting of 1,000,001 data points, it is shown that the runtime for performing RQA could be reduced from almost eight hours using a state-of-the-art software tool to roughly 69 seconds. Those dramatic performance improvements open the door for novel applications, such as multi-scale recurrence analysis~\citep{Sips2016}.

The development of \emph{PyRQA} is ongoing. The list of planned features include an improved separation of the analytical operators within the source code as well as an automatic selection of the best-performing RQA implementation at runtime. User feedback furthermore suggests to support the quantitative analysis of precomputed recurrence matrices.    

\section{Acknowledgements}
This work is supported by grants from the Deutsche Forschungsgemeinschaft, Graduiertenkolleg METRIK (GRK 1324) and Graduiertenkolleg ``Natural Hazards and Risks in a Changing World'' (GRK 2043/1).

\appendix

\section*{References}

\bibliography{mybibfile}

\end{document}